 \documentclass[final,5p,times,twocolumn]{elsarticle}

 \usepackage{graphics}

\usepackage{amssymb}

\journal{Physics Letters A}

\begin{document}

\begin{frontmatter}



\title{Boundary crisis and suppression of Fermi acceleration in a dissipative two
dimensional non-integrable time-dependent billiard.}


\author{Diego F. M. Oliveira}

\address{Departamento de F\'isica -- Instituto de Geoci\^encias e 
Ci\^encias Exatas -- Universidade Estadual Paulista -- Av. 24A, 1515 -- Bela Vista -- CEP:
13506-900 -- Rio Claro -- SP -- Brazil}

\ead{diegofregolente@gmail.com}

\author{Edson D.\ Leonel}

\address{Departamento de Estat\'{\i}stica, Matem\'atica Aplicada e
Computa\c c\~ao -- Instituto de Geoci\^encias e Ci\^encias Exatas --
Universidade Estadual Paulista -- Av. 24A, 1515 -- Bela Vista -- CEP:
13506-900 -- Rio Claro -- SP -- Brazil}

\ead{edleonel@rc.unesp.br}

\begin{abstract}
Some dynamical properties for a dissipative time-dependent oval-shaped
billiard are studied. The system is described in terms of a
four-dimensional nonlinear mapping. Dissipation is introduced via
inelastic collisions of the particle with the boundary, thus implying
that the particle has a fractional loss of energy upon collision. The
dissipation causes profound modifications in the dynamics of the
particle as well as in the phase space of the non dissipative system. In
particular, inelastic collisions can be assumed as an efficient
mechanism to suppress Fermi acceleration of the particle. The
dissipation also creates attractors in the system, including chaotic.
We show that a slightly modification of the intensity of the
damping coefficient yields a drastic and sudden destruction of the
chaotic attractor, thus leading the system to experience a boundary
crisis. We have characterized such a boundary crisis via a collision of
the chaotic attractor with its own basin of attraction and confirmed
that inelastic collisions do indeed suppress Fermi acceleration in
two-dimensional time dependent billiards.

\end{abstract}

\begin{keyword}
Billiard, Chaos, Boundary crisis.

\end{keyword}

\end{frontmatter}


\section{Introduction}
\label{sec1}
During the last decades many theoretical studies on dissipative systems
have been introduced in order to explain different physical phenomena
in different fields of science including atomic and molecular physics
\cite{ref1,ref2,ref3}, turbulent and fluid dynamics
\cite{ref4,ref5,ref6,ref7}, optics \cite{ref8,ref9,ref10},
nanotechnology \cite{ref11,ref12}, quantum and relativistic systems
\cite{ref13,ref14,ref15}. Different procedures have been used to
describe such systems and two main different approaches are: (i) solving
differential ordinary/partial equations or; (ii) using the so called
billiard formalism. In principle, to chose procedure (i) or (ii)
strongly depends on the type of system considered and possible
existing symmetries. Case (i) are more likely devoted to systems where
the external potential are smooth while case (ii) describes situations
where the potential is null, say inside the boundary, and infinity
outside of the boundary. The boundary identifies the position of this
abrupt change. In this paper we shall concentrate to study case (ii),
i.e. a billiard system.

A billiard consists of system in which one or many point-like particles
move freely inside a closed region suffering specular
reflections/collisions with the boundary. Billiards can be considered
one of the most attractive types of dynamical models in the study of
ergodic and mixed properties in Hamiltonian systems \cite{zasl}. From
the mathematical point of
view, a billiard is defined by a connected region $Q\subset R^D$, with
boundary $\partial Q\subset R^{D-1}$ which separates $Q$ from its
complement. If $\partial Q=\partial Q(t)$ the system has a
time-dependent boundary and it can exchange energy with the particle
upon collision. Moreover, dissipation can be considered via different
ways where the most common types used (i) drag force; (ii) damping
coefficients. In the first case the particle loses energy/velocity as it
were moving immersed in a fluid. For such a case, the dynamics is
described by solving differential equations \cite{ref15_a}. On the
other hand, in the case (ii), the particle loses energy/velocity upon
collision with the moving boundary. Thus the system is normally
described using a billiard approach. 
It is know that
depending on the combination of initial conditions and control
parameters, the phase space of such systems possess different
structures. In the absence of dissipation, one kind of structure is the
mixed type \cite{ref16,ref17,ref18,ref19,ref20,ref21,ref22,ref23} where
regular regions, such as invariant tori and
Kolmogorov-Arnold-Moser (KAM) islands are observed coexisting with
chaotic seas. It is also well known in the literature that depending on
the structure of the phase space the system can show or not a phenomenon
called as Fermi acceleration, i.e., unlimited energy growth \cite{ref23_aa}. Such a
nomenclature comes from Enrico Fermi \cite{ref23_a} in 1949, as an
attempt to explain the origin of cosmic ray acceleration. He proposed
that such phenomenon was due to the interaction between charged
particles and time-dependent magnetic structures in the space. Since
then the model has been modified and studied considering different
approaches. One of the most studied version of the problem is the
Fermi-Ulam model (FUM) \cite{ref23_2,ref23_3}. Such model consists of a classical point-like
particle moving between two rigid walls, one of them is assumed to be
fixed and the other one moves according to a periodic function. In such
system, Fermi acceleration is not observed since the phase space has a
set of invariant tori limiting the size of the chaotic sea.
However, an alternative model was proposed by Pustylnikov
\cite{ref23_b} which is often called as bouncer model. Such system consists of a
classical particle falling due to the action of a constant gravitational
field on a moving platform. One of the most important properties of this
system is that depending on the combinations of both initial conditions
and control parameters, the unlimited energy gain for a classical
particle can be observed. It happens because there is no invariant tori 
limiting the size of the chaotic sea.

When dissipation is taken into account, one can show that the mixed
structure of the phase space present in the conservative case is
destroyed. Then, an elliptic fixed point (generally surrounded by KAM
islands) turns into a sink. Regions of chaotic seas might be replaced by
chaotic attractors. Each one of these attractors has its own basin of
attraction. Then, as a slight increase on the value of the damping
coefficient, that  is equivalent to reduce the power of dissipation, the chaotic
attractor touch, even crosses, the line separating the basin of attraction of the
chaotic attractor and the attracting fixed point (sink). Such behaviour
yields in a sudden destruction of the chaotic attractor. This
destruction is called as a boundary crisis \cite{ref24,ref25,ref26}.
After the destruction, the chaotic attractor is replaced by a
chaotic transient and its basin of attraction is destroyed, too. Additionally, when dissipation is considered 
the behaviour of energy changes from unlimited to a constant plateau for long enough time. Thus, confirming that
the phenomenon of Fermi acceleration is suppressed by dissipation \cite{ref26_a,ref26_b}. 

In the present paper we are interested in characterizing a boundary crisis
in a time-dependent oval-shaped billiard. The paper is organized as
follows. In Sec. \ref{sec2} we describe all the need details to
obtain the four-dimensional mapping that describe the dynamics of the
system. Our numerical results are discussed in Sec. \ref{sec_nova}.
Conclusion and acknowledgments are drawn in Sec. \ref{sec3}.

\section{The model and the mapping}
\label{sec2}

\begin{figure}[t]
\centerline{\includegraphics[width=0.95\linewidth]{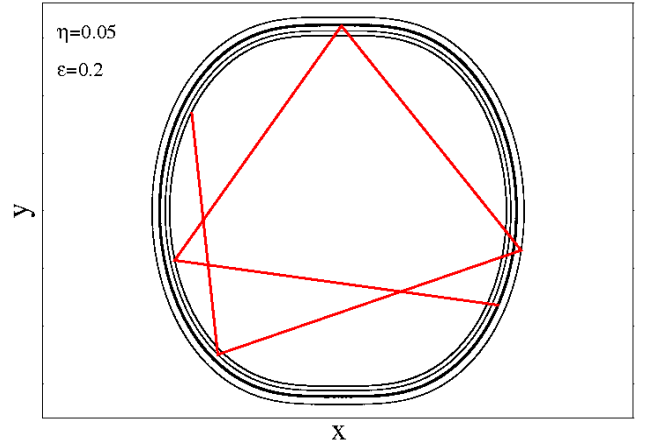}}
\caption{\it{ Plot of the boundary and a typical trajectory shown
only for 6 snapshots of the boundary. The control parameters used were
$\epsilon=0.2$, $\beta=\gamma=1.0$ and $\eta=0.05$.}}
\label{fig_new}
\end{figure}

The two dimensional time-dependent oval billiard consists of a classical
particle of mass {\it m} confined in and suffering collisions with a
periodically moving boundary. The model is 2-dimensional (2-D) in the
sense that it has two degrees of freedom, however, the dimension of the
phase space is defined as $2\times D$. Based on this fact, we described
the model using a four dimensional and non linear map
$T(\theta_{n},\alpha_{n},\overrightarrow{V}_{n},t_{n})=(\theta_{n+1},
\alpha_{n+1},\overrightarrow{V}_{n+1}
,t_{n+1})$ where the dynamical variables are, respectively, the angular
position of the particle; the angle that the trajectory of the particle
does with the tangent line at the position of the collision; the
absolute velocity of the particle; and the instant of the hit with the
boundary. The shape of the boundary is given in polar coordinates as
$
R_b(\theta,\epsilon,\eta,t)=[1+\eta\cos(t)][1+\epsilon\cos(2\theta)]
$
where $\epsilon$ is the circle's boundary deformation and $\eta$ is the
amplitude of the time dependent perturbation and the sub-index $b$
denotes boundary. Figure \ref{fig_new}
shows a typical plot of the boundary and six collisions of a particle
with the boundary.

To construct the mapping, we start with an initial condition
$(\theta_n,\alpha_n,V_n,t_n)$. The Cartesian components of
the boundary at the angular position $(\theta_n,t_n)$ are
\begin{eqnarray}
X(\theta_{n},t_n)&=&[1+\eta\cos(t_n)][1+\epsilon\cos(2\theta_n)]\cos(\theta_{n})~, \\ 
Y(\theta_{n},t_n)&=&[1+\eta\cos(t_n)][1+\epsilon\cos(2\theta_n)]\sin(\theta_{n})~.
\label{eq2}
\end{eqnarray}
The angle between the tangent of the boundary at the position
$(X(\theta_{n}),Y(\theta_{n}))$ measured with respect to the horizontal
line is
\begin{eqnarray}
\phi_n=\arctan\left[Y'(\theta_n,t_n) \over X'(\theta_n,t_n) \right]~,
\label{eq03}
\end{eqnarray}
where the expressions for both $X'(\theta_n,t_n)$ and $Y'(\theta_n,t_n)$
are written as
\begin{eqnarray}
X'(\theta_n,t_n)&=&{dR(\theta_n,t_n) \over d\theta_n} \cos(\theta_n)- R(\theta_n,t_n)\sin(\theta_n),\\
Y'(\theta_n,t_n)&=&{dR(\theta_n,t_n) \over d\theta_n} \sin(\theta_n)+
R(\theta_n,t_n)\cos(\theta_n)~,
\label{eq3}
\end{eqnarray}
with $dR(\theta_n,t_n)/d\theta_n=-2\epsilon[1+\eta
\cos(t_n)]\sin(2\theta_n)$. Since the expressions for $\phi_n$ and
$\alpha_n$ are known, the angle of the trajectory of the particle
measured with respect to the positive X-axis is $(\phi_n+\alpha_n)$.
Such information allows us to write the particle's velocity vector as
\begin{eqnarray}
\overrightarrow{V}_n=\vert\overrightarrow{V_n}\vert [\cos(\phi_n+\alpha_n)\widehat{i}+\sin(\phi_n+\alpha_n)\widehat{j}]~,
\label{eq4}
\end{eqnarray}
where $\widehat{i}$ and $\widehat{j}$ denote the unity vectors with
respect to the X and Y axis, respectively. The position of the particle,
as a function of time, for $t \geq t_n$, is given by
\begin{eqnarray}
X_p(t)&=&X(\theta_n,t_n)+\vert\overrightarrow{V}_n\vert\cos(\phi_n + \alpha_n)(t-t_n)~,\\
Y_p(t)&=&Y(\theta_n,t_n)+\vert\overrightarrow{V}_n\vert\sin(\phi_n + \alpha_n)(t-t_n)~.
\label{eq5}
\end{eqnarray}
We stress the sub-index $p$ denotes that such coordinates correspond to
the particle. The distance of the particle measured with respect to the
origin of the coordinate system is given by 
$R_p(t)=\sqrt{X^2_p(t)+Y^2_p(t)}$ and $\theta_p$ at ($X_p(t),Y_p(t)$) is $\theta_p=\arctan[Y_p(t)/X_p(t)]$. Therefore, the angular
position at the next collision of the particle with the boundary, i.e.
$\theta_{n+1}$, is numerically obtained by solving
$R_p(\theta_{n+1},t_{n+1})=R_b(\theta_{n+1},t_{n+1})$. It means that the
position of the boundary is the same as the position of the particle at
the instant of the collision. The time $t_{n+1}$ is obtained by
evaluating the expression 
\begin{eqnarray}
t_{n+1}=t_n+ {{\sqrt{\Delta X^2+\Delta Y^2}} \over \vert\overrightarrow{V}_n\vert}~,
\label{eq6}
\end{eqnarray}
where $\Delta X=X_p(\theta_{n+1},t_{n+1})-X(\theta_{n},t_{n})$ and
$\Delta Y=Y_p(\theta_{n+1},t_{n+1})-Y(\theta_{n},t_{n})$. To obtain
the new velocity we should note that the referential frame of the
boundary is moving. Since we are considering inelastic collisions, the
particle experiences a fractional loss of energy upon collision in both
its normal and tangential components. 
Therefore, at the instant of collision, the following conditions must be matched
\begin{eqnarray}
\overrightarrow{V^\prime}_{n+1}\cdot\overrightarrow{T}_{n+1}
&=&\beta\overrightarrow{V^\prime}_{n}\cdot\overrightarrow{T}_{n+1}~,\\
\overrightarrow{V^\prime}_{n+1}\cdot\overrightarrow{N}_{n+1}
&=&-\gamma\overrightarrow{V^\prime}_{n}\cdot\overrightarrow{N}_{n+1}~,
\label{eq7}
\end{eqnarray}
where the unitary tangent and normal vectors are
\begin{eqnarray}
\overrightarrow{T}_{n+1}&=&\cos(\phi_{n+1})\widehat{i}+\sin(\phi_{n+1})\widehat{j}~,\\
\overrightarrow{N}_{n+1}&=&-\sin(\phi_{n+1})\widehat{i}+\cos(\phi_{n+1})\widehat{j}~.
\label{eq8}
\end{eqnarray}
$\beta$ and $\gamma$ are damping coefficients, it means that the
particle can loses velocity/energy upon collision in its normal
component $(\gamma)$, tangential component $(\beta)$ or both. We
consider both $\gamma\in[0,1]$ and $\beta\in[0,1]$. The completely
inelastic collision happens when $\gamma=\beta=0$ and is not considered
in this paper. On the other hand, when $\gamma=\beta=1$, corresponding
to an elastic collision, all the results for the non-dissipative case
are recovered. The upper prime indicates that the velocity of the
particle is measured with respect to the moving boundary referential frame.
At the new angular position $\theta_{n+1}$,  we find that
\begin{eqnarray}
\overrightarrow{V}_{n+1}\cdot\overrightarrow{T}_{n+1}
&=&\beta\overrightarrow {V}_{n}\cdot\overrightarrow{T}_{n+1}
+\nonumber\\&+&(1-\beta)\overrightarrow{
V}_{b}(t_{n+1})\cdot\overrightarrow{T}_{n+1}~,
\label{eq8a}
\end{eqnarray}
\begin{eqnarray}
\overrightarrow{V}_{n+1}\cdot\overrightarrow{N}_{n+1}
&=&-\gamma\overrightarrow{V}_{n}\cdot\overrightarrow{N}_{n+1}
+\nonumber\\&+&(1+\gamma)\overrightarrow{V}_{b}(t_{n+1}
)\cdot\overrightarrow {N}_{n+1}~,
\label{eq8b}
\end{eqnarray}
where $\overrightarrow{V}_{b}(t_{n+1})$ is the velocity of the boundary which is written as
\begin{eqnarray}
\overrightarrow{V}_{b}(t_{n+1})={dR_{b}(t_{n+1}) \over dt_{n+1}} [\cos(\theta_{n+1})\widehat{i}+\sin(\theta_{n+1})\widehat{j}]~,
\label{eq10}
\end{eqnarray}
with
\begin{eqnarray}
{dR_{b}(t_{n+1}) \over dt_{n+1}}= -\eta[1+\epsilon\cos(2\theta_{n+1})]\sin(t_{n+1})~.
\label{eq11}
\end{eqnarray}

\begin{figure}[t]
\centerline{\includegraphics[width=0.95\linewidth]{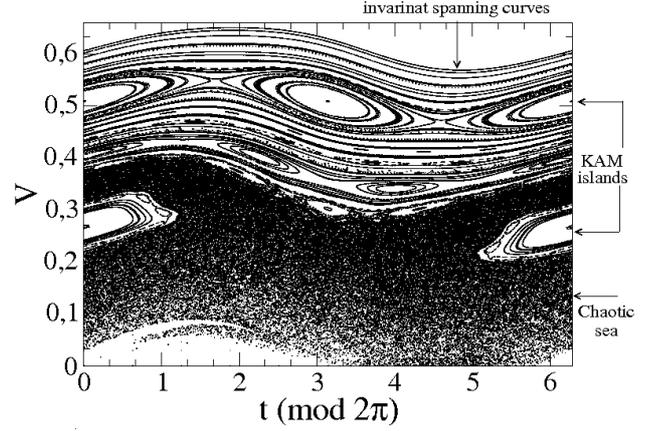}}
\caption{\it{Phase space in the variables velocity and time for a time
dependent oval billiard. The control parameters used were
$\epsilon=0.2$, $\beta=\gamma=1.0$ and $\eta=0.05$. We assume as fixed
the initial conditions $\alpha_0=\pi/2$ and $\theta_0=3\pi/2$.}}
\label{fig710}
\end{figure}

Then we have
\begin{eqnarray}
|\overrightarrow{V}_{n+1}|=\sqrt{(\overrightarrow{V}_{n+1}\cdot\overrightarrow{T}_{n+1}
)^2+(\overrightarrow{V}_{n+1}\cdot\overrightarrow{N}_{n+1})^2}~.
\label{eq012}
\end{eqnarray}

Finally, the angle $\alpha_{n+1}$ is written as
\begin{eqnarray}
\alpha_{n+1}=\arctan
\left[{\overrightarrow{V}_{n+1}\cdot\overrightarrow{N}_{n+1} \over
\overrightarrow{V}_{n+1}\cdot\overrightarrow{T}_{n+1}} \right]~.
\label{eq013}
\end{eqnarray}

With this four dimensional mapping, we can explore now numerical
results for the dynamics of the particle.

\subsection{Numerical Results}
\label{sec_nova}

In this section we discuss our numerical results. Just to remind, our
main goal is to characterize a boundary crisis in a time-dependent
oval-shaped billiard. To start, we show in Fig. \ref{fig710} a typical
phase space for a special set of initial conditions: $\alpha_0=\pi/2$
and $\theta_0=3\pi/2$. For such combination of initial condition and taken into account $\epsilon=0.2$ and $\eta=0.05$ the boundary has
neutral curvature. With this particular choice of initial
conditions, the phase space of the system is mixed. On the other hand,
if we chose random $\alpha_0$ and $\theta_0$, the particle experiences
the phenomenon of unlimited energy growth \cite{syok}.

\begin{figure}[t]
\vspace*{-0.8cm}
\centerline{\includegraphics[width=0.95\linewidth]{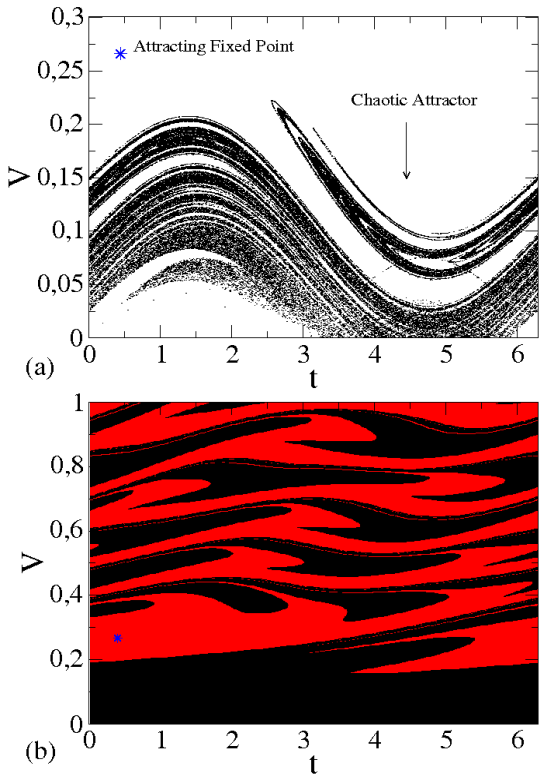}}
\caption{\it{(Color online) (a) Attracting Fixed point indicated by a
blue star ($\ast$) and chaotic attractor (set of black points); (b)
Their corresponding basin of attraction. Black corresponds to the basin
of attraction for the chaotic attractor and dark gray (red) of the
attracting fixed point. The control parameters used were $\epsilon=0.2$,
$\beta=0.25$, $\gamma=0.8899$ and $\eta=0.05$.}}
\label{fig711}
\end{figure}

\begin{figure}[t]
\vspace*{-0.8cm}
\centerline{\includegraphics[width=0.95\linewidth]{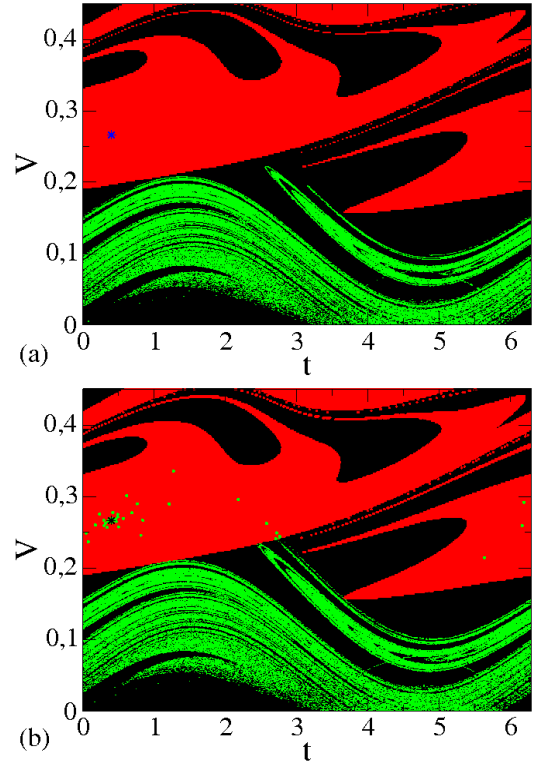}}
\caption{\it{(Color online) Basin of attraction for the chaotic
attractor and attracting fixed point (sink). The region in black corresponds to the basin of
attraction of the chaotic attractor; the region in dark gray (red),
denotes the basin of attraction of the attracting fixed point; light gray
(green) in (a) identifies the chaotic attractor. The control parameters used to
construct the basin of attraction were $\epsilon=0.2$, $\beta=0.25$,
$\eta=0.05$ and $\gamma=0.8899$. The dissipation used in (a) the chaotic attractor were $\gamma=0.8899$ (before crisis, light gray (green)); (b)
the chaotic transient were $\gamma=0.8906$ (after crisis, light gray (green)).}}
\label{fig712}
\end{figure}

We now consider the situation where both damping coefficients
$\beta\ne 1$ and $\gamma\ne 1$. We then keep fixed up to the end of the
paper $\beta=0.25$. It implies that there is a high dissipation along
the tangent component of the particle's velocity. Results for different
$\beta$ will be published elsewhere \cite{ref27_a}. The parameter $\gamma$ is
considered from the order of $\gamma=0.89$. It is shown in Fig.
\ref{fig711}(a) the behavior of the attractors present in the system
for the following combination of control parameters $\epsilon=0.2$,
$\beta=0.25$, $\gamma=0.8899$ and $\eta=0.05$. We can see a chaotic
attractor and an attracting fixed point. Figure \ref{fig711}(b) shows
their corresponding basin of attraction. The procedure used to construct
the basin of attraction was divide both $V$ and $t$ into windows of $500$
parts each, thus leading to a total of $2.5 \times 10^5$ different
initial conditions. Each initial condition was iterated up to $n=5
\times 10^6$ collisions with the boundary. We see that only two
attractors emerged for such combination of control parameters: sink and
chaotic attractor. We stress that other attractors could in principle
exist. If they exist however, their basin of attraction are too small to
be obtained. It is clear that, after a very long number of collisions of
the particle with the boundary, the velocity of the particle does not
grow unlimitedly. Consequently, no Fermi acceleration is observed and we
conclude that introduction of inelastic collisions worked out perfectly
as a mechanism to suppress Fermi acceleration, as proposed in Ref.
\cite{leonel_jpa} for a stochastic 1-D system.

Let us now go ahead with the characterization of the boundary crisis
\cite{ref24,ref25,ref26}. It is well known in the literature that a saddle
fixed point, in the plane $V\times t$ has two kinds of manifolds: (a)
stable and (b) unstable. The unstable manifolds are
formed by a family of trajectories that turn away from the saddle fixed
point. One of them can form the chaotic attractor (or visit the region
of the chaotic attractor after the event of crisis), while the other
one moves towards an attracting fixed point. These manifolds are
obtained from the iteration of the map $T$ with appropriate initial
conditions. Similarly, the construction of stable manifolds are a
little bit more complicated since the inverse of the mapping,
say $T^{-1}$, must be obtained. The procedure for obtaining the stable
manifolds is the same as that one used for the unstable manifolds,
however, instead of iterating the map $T$ we must iterate its
inverse $T^{-1}$. Since the stable manifolds generate the border of
the basin of attraction of the chaotic attractor and attracting sink, a
boundary crisis happens when a chaotic attractor touch the stable manifold due to a modification of the control parameter. As
a consequence, there is a sudden and drastic destruction of the chaotic
attractor and its basin of attraction.

\begin{figure}[t]
\centerline{\includegraphics[width=1.0\linewidth]{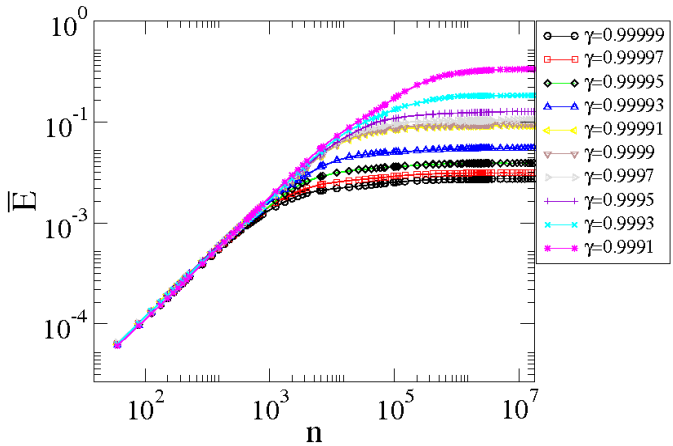}}
\caption{\it{Behaviour of $\overline{E} \times n$ for different values of $\gamma$, as labeled in the figure. The control parameters
used in the construction of the figure were $p=2$, $\epsilon=0.4$, $\beta=1$ and $\eta=0.001$.}}
\label{fig714}
\end{figure}

It is shown in Fig. \ref{fig712}(a), two basins of attraction; one in
black, corresponding to the basin of attraction of the chaotic
attractor, and the other one in dark gray (red), denoting the basin of
attraction of the attracting fixed point, and the chaotic attractor marked
by light gray (green). If we increase the value of the parameter
$\gamma$, which is equivalent to reduce the intensity of the
dissipation, the two branches of the stable manifold touch, even crosses, the edges
of the chaotic attractor, see Fig. \ref{fig712} (b). Such behaviour is
equivalent to a {\it collision} of the chaotic attractor with its own
basin of attraction. Of course after the {\it collision}, there is a
sudden destruction of chaotic attractor and its basin of attraction.
We have observed such crisis for the control parameters $\epsilon=0.2$,
$\beta=0.25$, $\eta=0.05$ and $\gamma=0.8906$. Other boundary crisis
are observed for different combinations of control parameters, too. After
the boundary crisis, the entire plane $V\times t$ (t mod ($2\pi$))
is the basin of attraction for a sink. Therefore, all initial conditions
in such a region will converge to the sink. Additionally, for the regime of weak dissipation, 
the average energy, $\overline{E}_i={{1}\over{n+1}}\sum_{j=0}^nE_{i,j}$, grows for short iteration number and suddenly it bends towards a regime of
saturation for long enough values of $n$ as can be seen in Fig. \ref{fig714}. Consequently, the mechanism of Fermi
acceleration is suppressed in high as well as weak dissipation.

\section{Conclusion}
\label{sec3}

As a short remark, we have studied a classical version of a dissipative
time-dependent oval-shaped billiard. The dissipation was introduced via
damping coefficients for both the normal and tangential components
of the particle's velocity. For the regime of high tangential
dissipation, we characterized an event of boundary crisis. For the
regime of weak dissipation, we have shown that the average energy
remains constant for long enough time. Such result allows us to
confirm that the introduction of inelastic collisions is sufficient
to suppress Fermi acceleration since all the initial conditions will
converge to attractors located at low velocity domain.

\section*{ACKNOWLEDGMENTS}
D.F.M.O gratefully acknowledges FAPESP. E. D. L. is grateful to FAPESP,
CNPq and FUNDUNESP, Brazilian agencies. The authors acknowledge Dr. J\"urgen
Vollmer for a careful reading on the manuscript.

\end{document}